# Galactic cosmic ray induced radiation dose on terrestrial exoplanets

Dimitra Atri[1,2], B. Hariharan[2], and Jean-Mathias Grießmeier[3,4]

1. Blue Marble Space Institute of Science, Seattle, WA 98145-1561, USA,

    Email: dimitra@bmsis.org

2. Tata Institute of Fundamental Research, Mumbai 400 005, India

3. Laboratoire de Physique et Chimie de l'Environnement et de l'Espace,

LPC2E CNRS/Universite d'Orleans, 45071 Orleans Cedex 02, France

4. Station de Radioastronomie de Nancay, Observatoire de Paris, CNRS/INSU,

18330 Nancay, France.

This past decade has seen tremendous advancements in the study of extrasolar planets. Observations are now made with increasing sophistication from both ground and space based instruments, and exoplanets are characterized with increasing precision. There is a class of particularly interesting exoplanets, falling in the habitable zone, which is defined as the area around a star where the planet is capable of supporting liquid water on its surface. Planetary systems around M dwarfs are considered to be prime candidates to search for life beyond the solar system. Such planets are likely to be tidally locked and have close-in habitable zones. Theoretical calculations also suggest that close-in exoplanets are more likely to have weaker planetary magnetic fields, especially in case of super earths. Such exoplanets are subjected to a high flux of Galactic Cosmic Rays (GCRs) due to their weak magnetic moments. GCRs are energetic particles of astrophysical origin, which strike the planetary atmosphere and produce secondary particles, including muons, which are highly penetrating. Some of these particles reach the planetary surface and contribute to the radiation dose. Along with the magnetic field, another factor governing the radiation dose is the depth of the planetary atmosphere. The higher the depth of the planetary atmosphere, the lower the flux of secondary particles will be on the surface. If the secondary particles are energetic enough, and their flux is sufficiently high, the radiation from muons can also impact the sub-surface regions, such as in the case of Mars. If the radiation dose is too high, the chances of sustaining a long-term biosphere on the planet are very low. We explore the dependence of the GCR induced radiation dose on the strength of the planetary magnetic field and its atmospheric depth, finding that the latter is the decisive factor for the protection of a planetary biosphere.

## 1. INTRODUCTION

What are the physical conditions that make a planet habitable? The solution to this problem depends on the definition of habitability. One way to approach this problem is to study the Earth and estimate the range of physical conditions which can support an Earth-like biosphere. These include the astrophysical conditions such as the stellar spectrum and flux and also the properties of the planetary atmosphere for climate modeling. There is a tremendous interest in the search for signatures of life on planets around stellar systems, which can support liquid water on its surface (Kasting et al., 1993). However, here we focus on a different approach, where we estimate the range of physical conditions for which the radiation dose can permit a stable Earth-like biosphere. We explore various physical conditions that give rise to increased radiation dose on an exoplanet's surface. The radiation environment of a planet consists not only of the photon and proton flux from the host star, but also the galactic cosmic ray (GCR) flux consisting of charged nuclei (mostly protons). Although the flux of GCRs is only a small fraction of the radiation flux from the host star, the average energy of individual GCR particles is higher by several orders of magnitude than photons and protons from the host star. The GCR flux depends on (1) the magnetic moment of the planet, and (2) the location of the planetary system at a particular time in the galaxy. GCR secondary particles comprise of the most penetrating ionizing radiation and its biological effects have been discussed extensively (Atri and Melott, 2013; Melott and Thomas, 2011; Dartnell, 2011).

While the atmospheric damage resulting from high-energy primary photons has been modeled extensively (Gehrels et al., 2003; Thomas et al., 2005; Ejzak et al., 2007), the atmospheric effects of GCRs have only been modeled partially using approximate analytical methods (Grenfell et al., 2007; Grenfell et al., 2012). This is a major shortfall in current studies, because GCR impact is much higher on planets with low magnetic moments (Grießmeier et al., 2005), therefore the currently used simple approach leads to a large error in calculating the concentrations of biomarker molecules. GCR primaries

(mostly protons) interact with the atmosphere producing secondary particles, also known as air showers. This shower comprises of the electromagnetic component, and the secondary component containing charged particles, which propagate towards the planetary surface along with the shower (Gaisser, 1991). The electromagnetic component ionizes the atmosphere, which can significantly alter the atmospheric chemistry in the upper atmosphere (Nicolet, 1975; Thomas et al., 2005).

The secondary component primarily consists of muons, neutrons and electrons. The most energetic of them are primarily muons, and depending on their energy, they can even penetrate several hundred meters below the planetary surface (Gaisser, 1991). Increase in muon flux can have serious biological implications such as increase in the mutation rate and DNA damage (Dar et al., 1998) for both terrestrial and marine life. Other particles such as electrons and neutrons also produce various kinds of biological damage (Alpen, 1997). Under certain physical conditions, there could be a significantly higher flux of secondary particles and the resulting biological radiation dose should be considered as an important factor in constraining the habitability of a planet. Planetary systems around M dwarfs are considered to be prime targets to search for life beyond the solar system. They are favorable because they are abundant in the Galaxy (Tarter et al., 2007; Scalo et al., 2007), provide a long term stable environment after the first 0.5 - 1 Gyr, and have close-in habitable zones, which are good for transit observations of potential habitable planets (Irvin et al., 2008). Theoretical arguments suggest that planets in the habitable zone around M dwarfs and close-in super earths have weak magnetic moments and would have a higher flux of GCRs (Khodachenko et al., 2011). With increasing sophistication in observational techniques, it will soon be possible to obtain transit spectra of atmospheres of several potential habitable planets (Gardner et al., 2006; Seager et al., 2009). Interpreting the observational data will require detailed photochemical modeling of the planetary atmosphere with its radiation environment (Segura et al., 2005). The radiation environment in turn can also potentially provide a constraint on the habitability of the planet.

## 2. METHODOLOGY

Since the GCR spectrum depends on the magnetic field parameters, we use a theoretical model to calculate the particle spectra of close-in terrestrial exoplanets. Then we use the spectra to propagate GCR particles with different depths of planetary atmospheres. We obtain the flux of secondary particles on the surface of the planet in each case, calculate the biological radiation dose and discuss its implications for a long-term sustained biosphere on that planet.

### Particle Spectra

The GCR flux reaching the top of the planetary atmosphere depends strongly on the planetary magnetic moment. Extrasolar planets orbiting in habitable zones around M dwarfs are tidally locked (Kasting et al., 1993). Theoretical arguments indicate that such planets are likely to have a weak magnetic moment, and would accordingly not be protected by an extended magnetosphere (Grießmeier et al., 2005, Grießmeier et al., in preparation). Stellar wind velocity pushes the planetary magnetosphere further, allowing more GCR particles into the planetary atmosphere. An estimate of their orbital magnetic moment can be made using a number of scaling laws, described in detail elsewhere (Grießmeier et al., 2005; Grießmeier et al., 2009).

A planet with weak magnetic moment would allow a larger flux of GCRs over a larger area compared to a strongly magnetized planet, such as the Earth. GCR flux then can be calculated as a function of the planetary magnetic moment. In order to evaluate the number of particles penetrating through a planetary magnetosphere, we have to select an appropriate magnetospheric model. In this work, the magnetosphere is assumed to be closed (i.e. magnetic field lines cannot cross the magnetopause), and is modeled as a cylinder (on the night side) topped by a hemisphere (on the day side). The radius of the hemisphere and the cylinder is determined by the pressure balance between the stellar wind ram pressure and the magnetic pressure of the planetary magnetic field (which is assumed to be a zonal dipole, see below).

Within the hemisphere, the magnetospheric magnetic field is described by a series of spherical harmonics, and within the tail a series of Bessel functions are used. This allows the model to take into account not only the intrinsic field of the planet, but also the magnetic fields created by the magnetopause currents. This magnetospheric model was originally developed by Voigt (Voight, 1981), and extended by Stadelmann et al. (Stadelmann et al., 2010), where the details of the model are described.

This model has already been applied to the case of extrasolar planets by Grießmeier et al. (Grießmeier et al., 2005; Grießmeier et al., 2009) and Grenfell et al. (Grenfell et al., 2007). As in the present case, the planetary magnetic field was assumed to be a zonal dipole. However, in those studies the magnetic dipole strength was estimated using simple scaling arguments. Here, we take a different approach, and vary the planetary magnetic dipole field between 0% and 300% of the terrestrial value. Thus, rather than applying a model for the planetary magnetic moment, we show how magnetic protection varies as a function of the planetary magnetic dipole moment. Note that for close-in exoplanets, and for the case of super-Earths, a magnetic moment smaller than that on Earth should be expected (Grießmeier et al., in preparation), making our result especially relevant for these classes of planets.

In order to quantify the protection of extrasolar Earth-like planets against galactic cosmic ray protons, we investigate the motion of galactic cosmic protons through the planetary magnetic field described above. As no solution in closed form exists, this type of study is only possible through the numerical integration of many individual trajectories (Smart et al. 2000).

In this work, we analyze 4 different magnetospheric configurations (i.e. field strength of the planetary magnetic dipole), and for each we look at 14 different energy cases, ranging from 64 MeV to 524 GeV. For each case, we numerically follow the trajectories of 28 million particles, corresponding to protons with different starting positions and starting velocity directions.

The particles are launched from the surface of a sphere (the center of the sphere coincides with the center of the planet) with a sufficiently large radius, making sure that all particles are launched outside the magnetosphere (except for those arriving from the tailward direction). As usual in cosmic ray tracing, the computing-intensive part is not the calculation of the particle trajectories, but the evaluation of a complex magnetic field for each particle position. For a specific case, Smart et al. (2000) estimate that the magnetic field calculation takes 90% of the total CPU time, and only 10% of the CPU time is used for the calculation of the particle's motion.

As soon as the particle enters the magnetosphere (the grey area in Figure 1), its motion is influenced by the planetary magnetic field. The trajectories are calculated using the numerical leapfrog method. The example of Figure 1 clearly shows two populations of particles: (a) Particles that are deflected by the magnetospheric magnetic field, and (b) Particles (mostly those close to the polar cusp) which manage to penetrate deep into the magnetosphere and are able to reach the top of the atmosphere. For each energy, we count the fraction of particles which reach the top of the planetary atmosphere (described by a spherical shell one hundred kilometers above the planetary surface). This allows us to calculate the energy spectrum. More details on the numerical calculation of the cosmic rays trajectories can be found here (Stadelmann et al., 2010).

A similar model has already been applied to the case of extrasolar planets by Grießmeier et al. (Grießmeier et al., 2005; Grießmeier et al., 2009) and Grenfell et al. (Grenfell et al., 2007; Grenfell et al., 2012). The main differences with respect to these previous studies are the following:

- As described above, the planetary magnetic moment is not assumed from a physical model, but used a free parameter, and the resulting energy spectrum is calculated for five different values of the planetary magnetic dipole.

- We have included the case of high-energy cosmic ray particles. Where the previous calculations were limited to the energy range 64 MeV to 8.2 GeV, we now calculate particles from 64 MeV to 524 GeV.

- The calculation of high-energy particles made it necessary to multiply the number of particles by a factor of 4 to reach a satisfying statistics.

More details on these calculations will be given in Grießmeier et al. (in preparation). Biological implications of low-energy particles (< 8 GeV) were already discussed by Grießmeier et al. (2005). In the following, we will discuss the capacity of high-energy particles (< 524 GeV) to generate secondary muons, which have a significant biological relevance.

**Air showers**

As described earlier, air showers are produced when GCR particles strike the Earth's atmosphere. In order to model the interaction of GCR particles with the planetary atmosphere, we will use CORSIKA (COsmic Ray SImulations for KAscade), which is a widely used Monte Carlo tool to model cosmic ray induced air showers from primaries in a wide energy range (Heck et al., 1998). The code is regarded as a gold standard in simulating the propagation of GCRs in the atmosphere (Risse et al, 2001; Bernlohr, 2000; Nagano et al., 2000; . The model is continuously tested against data from a number of experiments around the globe and updated frequently with new physics results. Simulations were carried out using CORSIKA v6990, a stable version of the code with updated interaction models. The code has already been demonstrated to reproduce air shower data with high accuracy (Atri et al., 2010; Atri and Melott, 2011; Overholt et al., 2013).

The CORSIKA package has a choice of eight hadronic interaction models (Heck et al., 1998), and appropriate models can be chosen depending on the energy range of the primaries and focus of the study. A total of 20 million proton primaries were generated using the SIBYLL model for high-energy hadronic interactions and GHEISHA model for low energy hadronic interactions for each case. At these energies, any combination of model can be chosen for this work, because all models are well calibrated to low energy particle interactions (1 GeV - 10 TeV). Particles with energy greater than 80 GeV are treated with the high-energy model and the rest with the low energy model. In the standard options of the code, the Earth's atmosphere is assumed to be a flat disc, which can give inaccurate results in this case. The CURVED option is therefore used to simulate particles falling at zenith angles above 70 degrees. The UPWARD option was used to treat the upward travelling particles. The input particle spectrum was obtained from the magnetospheric model described above. Only four representative cases were chosen to estimate the extreme range of radiation doses. The showers were simulated in the energy range 8 GeV to 0.25 TeV over the entire spectrum as shown in the figure 2. The simulations covered 4 different magnetosphere models to model earth and exoplanets (i.e., outside magnetosphere, 15%, 50%, 100% of magnetic moment) and 5 values atmospheric depths (i.e., 100 gcm−2 , 200 gcm−2 , 500 gcm−2 , 700 gcm−2, 1036 gcm−2 ). The energy cut-off set to the lowest possible values for the secondary particles (i.e. Hadron = 50 MeV, Muon = 10 MeV, e+- = 50 keV, gamma = 50 keV). For each shower, the total number of detected particles and their energy deposited were calculated at every 20 gcm−2 of atmospheric depth. The energy spectra of secondary particles were also calculated, since higher energy particles penetrate much deeper. Neutrons below 50 MeV were calculated using the cosmic ray neutron lookup table (Overholt et al., 2013).

# 3. RESULTS

The all particle flux obtained from simulations is calculated for different spectra from magnetic field moments, and different atmospheric thicknesses. The results are presented with 4 magnetic field parameters and 5 atmospheric depths. Other than the particle flux, we also present the energy distribution of particles. This is important because biological damage is proportional to the particle energy for some particles.

All values shown below are generated with 20 million primary particles. Flux is defined as the total number of particles reaching the ground level from 20 million primaries.

**TABLE I: All particle flux**

| Magnetic moment (%) | 100 gcm$^{-2}$ | 200 gcm$^{-2}$ | 500 gcm$^{-2}$ | 700 gcm$^{-2}$ | 1036 gcm$^{-2}$ |
|---|---|---|---|---|---|
| 0 | 2.60 × 10$^7$ | 1.45 × 10$^7$ | 1.19 × 10$^6$ | 2.59 × 10$^5$ | 2.83 × 10$^4$ |
| 15 | 2.27 × 10$^7$ | 1.37 × 10$^7$ | 1.20 × 10$^6$ | 2.64 × 10$^5$ | 2.11 × 10$^4$ |
| 50 | 1.08 × 10$^7$ | 7.04 × 10$^6$ | 7.18 × 10$^5$ | 1.65 × 10$^5$ | 1.97 × 10$^4$ |
| 100 | 6.96 × 10$^6$ | 4.70 × 10$^6$ | 5.04 × 10$^5$ | 1.16 × 10$^5$ | 1.51 × 10$^4$ |

It is well known that the largest number of secondary particles reaching the surface are electrons. Electrons are produced by charged particles ionizing the atmosphere, or by decay of unstable particles such as pions. Because of their low charge to mass ratio, they lose energy rapidly by radiation. The biological effects of electron

### TABLE II: Photon flux

| Magnetic moment (%) | 100 gcm$^{-2}$ | 200 gcm$^{-2}$ | 500 gcm$^{-2}$ | 700 gcm$^{-2}$ | 1036 gcm$^{-2}$ |
|---|---|---|---|---|---|
| 0 | 1.89 × 10$^7$ | 1.12 × 10$^7$ | 8.23 × 10$^5$ | 1.36 × 10$^5$ | 1.96 × 10$^4$ |
| 15 | 1.73 × 10$^7$ | 1.09 × 10$^7$ | 8.54 × 10$^5$ | 1.49 × 10$^5$ | 1.32 × 10$^4$ |
| 50 | 8.38 × 10$^6$ | 5.71 × 10$^6$ | 5.41 × 10$^5$ | 1.06 × 10$^5$ | 1.33 × 10$^4$ |
| 100 | 5.49 × 10$^6$ | 3.85 × 10$^6$ | 3.84 × 10$^5$ | 7.67 × 10$^4$ | 1.01 × 10$^4$ |

### TABLE III: Electron flux

| Magnetic moment (%) | 100 gcm$^{-2}$ | 200 gcm$^{-2}$ | 500 gcm$^{-2}$ | 700 gcm$^{-2}$ | 1036 gcm$^{-2}$ |
|---|---|---|---|---|---|
| 0 | 1.81 × 10$^6$ | 9.32 × 10$^5$ | 6.31 × 10$^4$ | 1.05 × 10$^4$ | 1.97 × 10$^3$ |
| 15 | 1.68 × 10$^6$ | 9.19 × 10$^5$ | 6.61 × 10$^4$ | 1.21 × 10$^4$ | 1.35 × 10$^3$ |
| 50 | 8.27 × 10$^5$ | 4.94 × 10$^5$ | 4.18 × 10$^4$ | 8.80 × 10$^3$ | 1.18 × 10$^3$ |
| 100 | 5.47 × 10$^5$ | 3.35 × 10$^5$ | 3.07 × 10$^4$ | 6.16 × 10$^3$ | 9.30 × 10$^2$ |

### TABLE IV: Muon flux

| Magnetic moment (%) | 100 gcm$^{-2}$ | 200 gcm$^{-2}$ | 500 gcm$^{-2}$ | 700 gcm$^{-2}$ | 1036 gcm$^{-2}$ |
|---|---|---|---|---|---|
| 0 | 2.35 × 10$^5$ | 1.56 × 10$^5$ | 2.85 × 10$^4$ | 1.18 × 10$^4$ | 4.77 × 10$^3$ |
| 15 | 2.26 × 10$^5$ | 1.59 × 10$^5$ | 3.00 × 10$^4$ | 1.28 × 10$^4$ | 4.40 × 10$^3$ |
| 50 | 1.17 × 10$^5$ | 8.88 × 10$^4$ | 2.16 × 10$^4$ | 9.99 × 10$^3$ | 4.34 × 10$^3$ |
| 100 | 7.82 × 10$^4$ | 6.05 × 10$^4$ | 1.61 × 10$^4$ | 7.60 × 10$^3$ | 3.46 × 10$^3$ |

exposure are limited because they have very low energy and small penetrating power, and can only cause damage to the superficial layers for most organisms. However, for thin bacterial films, they can be lethal. For example, a thick skinned or a marine organism is immune to such electrons. The energy cutoff for electrons was set to 50 keV, the lowest possible value in CORSIKA and also where the energy is high enough to cause minor damage for terrestrial organisms. It should be noted that even higher energy electrons are not capable of causing any biological damage to benthic marine life, because they will be stopped by the column of water above them.

Neutrons are very damaging and can contribute significantly to the radiation dose in the upper atmosphere. Since neutrons are electrically neutral, they cannot cause damage by ionization like other particles. They can collide with the nuclei and transfer some kinetic energy without causing much damage. Or, they can be absorbed by the nuclei, making them unstable and resulting in a gamma-ray emission. Neutrons are produced in large numbers, especially at higher altitudes and can pose health risks to airline crew. The number of neutrons is reduced at the ground, and they do not contribute significantly to the overall radiation dose from cosmic rays. The quality factor of neutrons depend on their energy and so their energy distribution is used here to calculate the biological radiation dose. As shown in the table, the number of neutrons above 50 MeV decrease significantly as we move towards the lower part of the atmosphere. This is because neutrons lose energy by multiple collisions and their energy goes down as they move lower in the atmosphere. Low energy neutrons form the significant population of the total number of neutrons in the lower atmosphere. As seen in the table, the total number of neutrons increase dramatically if low-energy neutrons are considered.

**TABLE V: Neutron flux above 50 MeV**

| Magnetic moment (%) | 100 gcm$^{-2}$ | 200 gcm$^{-2}$ | 500 gcm$^{-2}$ | 700 gcm$^{-2}$ | 1036 gcm$^{-2}$ |
|---|---|---|---|---|---|
| 0 | 2.38 × 10$^6$ | 6.27 × 10$^5$ | 2.08 × 10$^4$ | 2.15 × 10$^3$ | 7.75 × 10$^1$ |
| 15 | 1.78 × 10$^6$ | 5.51 × 10$^5$ | 2.06 × 10$^4$ | 2.39 × 10$^3$ | 3.30 × 10$^1$ |
| 50 | 7.17 × 10$^5$ | 2.37 × 10$^5$ | 1.10 × 10$^4$ | 1.58 × 10$^3$ | 9.34 × 10$^1$ |
| 100 | 4.40 × 10$^5$ | 1.53 × 10$^5$ | 7.54 × 10$^3$ | 1.13 × 10$^3$ | 4.00 × 10$^1$ |

**TABLE VI: Total neutron flux**

| Magnetic moment (%) | 100 gcm$^{-2}$ | 200 gcm$^{-2}$ | 500 gcm$^{-2}$ | 700 gcm$^{-2}$ | 1036 gcm$^{-2}$ |
|---|---|---|---|---|---|
| 0 | 5.03 × 10$^6$ | 2.21 × 10$^6$ | 2.73 × 10$^5$ | 1.01 × 10$^5$ | 2.08 × 10$^3$ |
| 15 | 3.53 × 10$^6$ | 1.73 × 10$^6$ | 2.50 × 10$^5$ | 9.03 × 10$^4$ | 1.97 × 10$^3$ |
| 50 | 1.44 × 10$^6$ | 7.44 × 10$^5$ | 1.14 × 10$^5$ | 4.00 × 10$^4$ | 9.03 × 10$^2$ |
| 100 | 8.41 × 10$^5$ | 4.50 × 10$^5$ | 7.30 × 10$^4$ | 2.55 × 10$^4$ | 5.61 × 10$^2$ |

Muons are produced in large number in the upper atmosphere and have much higher energy compared to any other particle reaching the ground. This is because of their small interaction cross section and relatively high decay constant. They lose only ~2 MeV per gcm$^{-2}$ of the atmosphere and rest of the energy is dissipated on the surface (Gaisser, 1991). A large number of muons, as a result reach the surface level. Below the surface, only the flux of muons is important since rest of the components can be easily blocked by a small water column. They pose the greatest threat to both terrestrial and marine organisms. The energy cutoff for muons was set to the lowest possible value of

10 MeV. Since muons are the most energetic secondary particles, below 10 MeV, the number of muons at the surface are negligible and therefore, not taken into consideration.

We start with normalising the total number of particles to the Earth value, i.e. 1036 gcm$^{-2}$ atmosphere and 100% magnetic moment. Rest of the values are scaled accordingly and give flux per unit area per time. The energy deposition of different particles is then calculated using available data in literature. The energy deposited is then displayed in units of J m$^{-2}$ Sec$^{-1}$. Here we perform calculations to compare different cases of radiation flux and will subsequently calculate the radiation dose. The biological damage is roughly proportional to the amount of energy deposited by the radiation in a biological sample. The radiation dose is defined as the energy deposited per unit mass of a substance. The SI unit of effective biological radiation dose is Sievert, and is defined as the product of the radiation dose and the quality factor of the radiation and the organ in consideration: $D = dE/dM \times Q \times W$. The radiation dose for each component was obtained for a sample object. We define the sample object as a 15 cm cube of water, which is a standard practice in radiation biophysics to access radiation impact on humans. The radiation dose will be computed using well established quality factors from literature (United States Nuclear Regulatory Commission, 2013).

For muons in the energy range considered here, the energy deposition is approximately 2 MeV per gcm$^{-2}$ (Beringer et al., 2012). For electrons, we use results from a Geant4 based simulation which provides us with energy dependent particle energy deposition (Francis et al., 2011). For neutrons, we use well established dose calculations at Fermilab (Cossairt, 2009; United States Nuclear Regulatory Commission,

2013) (Beringer et al., 2012). Stopping power of photons was also calculated using data provided in the particle data book (Beringer et al., 2012).

**TABLE VII: Energy deposition rate from muons in J m$^{-2}$ sec$^{-1}$**

| Magnetic moment (%) | 100 gcm$^{-2}$ | 200 gcm$^{-2}$ | 500 gcm$^{-2}$ | 700 gcm$^{-2}$ | 1036 gcm$^{-2}$ |
|---|---|---|---|---|---|
| 0 | 4.14 × 10$^{-8}$ | 2.74 × 10$^{-8}$ | 5.01 × 10$^{-9}$ | 2.08 × 10$^{-9}$ | 8.39 × 10$^{-10}$ |
| 15 | 3.97 × 10$^{-8}$ | 2.79 × 10$^{-8}$ | 5.28 × 10$^{-9}$ | 2.25 × 10$^{-9}$ | 7.75 × 10$^{-10}$ |
| 50 | 2.06 × 10$^{-8}$ | 1.56 × 10$^{-8}$ | 3.81 × 10$^{-9}$ | 1.76 × 10$^{-9}$ | 7.63 × 10$^{-10}$ |
| 100 | 1.38 × 10$^{-8}$ | 1.06 × 10$^{-8}$ | 2.83 × 10$^{-9}$ | 1.34 × 10$^{-9}$ | 6.10 × 10$^{-10}$ |

**TABLE VIII: Energy deposition rate from electrons in J m$^{-2}$ sec$^{-1}$**

| Magnetic moment (%) | 100 gcm$^{-2}$ | 200 gcm$^{-2}$ | 500 gcm$^{-2}$ | 700 gcm$^{-2}$ | 1036 gcm$^{-2}$ |
|---|---|---|---|---|---|
| 0 | 3.18 × 10$^{-7}$ | 1.64 × 10$^{-7}$ | 1.11 × 10$^{-8}$ | 1.85 × 10$^{-9}$ | 3.47 × 10$^{-10}$ |
| 15 | 2.96 × 10$^{-7}$ | 1.62 × 10$^{-7}$ | 1.16 × 10$^{-8}$ | 2.13 × 10$^{-9}$ | 2.38 × 10$^{-10}$ |
| 50 | 1.46 × 10$^{-7}$ | 8.70 × 10$^{-8}$ | 7.36 × 10$^{-9}$ | 1.55 × 10$^{-9}$ | 2.07 × 10$^{-10}$ |
| 100 | 9.63 × 10$^{-8}$ | 5.89 × 10$^{-8}$ | 5.40 × 10$^{-9}$ | 1.08 × 10$^{-9}$ | 1.64 × 10$^{-10}$ |

**TABLE IX: Energy deposition rate from photons in J m$^{-2}$ sec$^{-1}$**

| Magnetic moment (%) | 100 gcm$^{-2}$ | 200 gcm$^{-2}$ | 500 gcm$^{-2}$ | 700 gcm$^{-2}$ | 1036 gcm$^{-2}$ |
|---|---|---|---|---|---|
| 0 | 1.93 × 10$^{-6}$ | 9.66 × 10$^{-7}$ | 6.28 × 10$^{-8}$ | 1.01 × 10$^{-8}$ | 1.91 × 10$^{-9}$ |
| 15 | 1.82 × 10$^{-6}$ | 9.68 × 10$^{-7}$ | 6.67 × 10$^{-8}$ | 1.13 × 10$^{-8}$ | 1.15 × 10$^{-9}$ |
| 50 | 9.45 × 10⁻⁷ | 5.33 × 10⁻⁷ | 4.42 × 10⁻⁸ | 8.49 × 10⁻⁹ | 1.03 × 10⁻⁹ |
| 100 | 6.35 × 10⁻⁷ | 3.69 × 10⁻⁷ | 3.20 × 10⁻⁸ | 6.16 × 10⁻⁹ | 8.42 × 10⁻¹⁰ |

**TABLE X: Energy deposition rate from neutrons in J m⁻² sec⁻¹**

| Magnetic moment (%) | 100 gcm⁻² | 200 gcm⁻² | 500 gcm⁻² | 700 gcm⁻² | 1036 gcm⁻² |
|---|---|---|---|---|---|
| 0 | 3.45 × 10⁻⁷ | 9.09 × 10⁻⁸ | 2.95 × 10⁻⁹ | 3.50 × 10⁻¹⁰ | 4.04 × 10⁻¹² |
| 15 | 2.73 × 10⁻⁷ | 8.14 × 10⁻⁸ | 2.94 × 10⁻⁹ | 3.81 × 10⁻¹⁰ | 1.38 × 10⁻¹¹ |
| 50 | 1.13 × 10⁻⁷ | 3.61 × 10⁻⁸ | 1.61 × 10⁻⁹ | 2.38 × 10⁻¹⁰ | 1.06 × 10⁻¹¹ |
| 100 | 7.05 × 10⁻⁸ | 2.34 × 10⁻⁸ | 1.11 × 10⁻⁹ | 1.70 × 10⁻¹⁰ | 5.43 × 10⁻¹² |

Now we calculate the effective biological radiation dose for our test object from individual radiation types.

**TABLE XI: Radiation dose from muons in mSv yr⁻¹**

| Magnetic moment (%) | 100 gcm⁻² | 200 gcm⁻² | 500 gcm⁻² | 700 gcm⁻² | 1036 gcm⁻² |
|---|---|---|---|---|---|
| 0 | 8.71 | 5.77 | 1.05 | 0.44 | 0.18 |
| 15 | 8.35 | 5.87 | 1.11 | 0.47 | 0.16 |
| 50 | 4.33 | 3.28 | 0.80 | 0.37 | 0.16 |
| 100 | 2.89 | 2.24 | 0.60 | 0.28 | 0.13 |

## 4. DISCUSSION

We have studied the surface radiation dose on terrestrial exoplanets with varying magnetic moments and atmospheric thickness and used the Earth's atmosphere to calculate the particles fluxes in all cases. All hadronic interactions depend on the average atomic mass, which does not change much depending on the atmospheric composition. Planetary exploration in the solar system shows most planetary atmospheric compositions consist of different percentages of C, N, O elements, which have similar atomic masses. Different atmospheric compositions might give different results by only a few percent.

**TABLE XII: Radiation dose from electrons in mSv yr$^{-1}$**

| Magnetic moment (%) | 100 gcm$^{-2}$ | 200 gcm$^{-2}$ | 500 gcm$^{-2}$ | 700 gcm$^{-2}$ | 1036 gcm$^{-2}$ |
|---|---|---|---|---|---|
| 0 | 66.87 | 34.49 | 2.34 | 0.39 | 0.07 |
| 15 | 62.18 | 33.99 | 2.45 | 0.45 | 0.05 |
| 50 | 30.61 | 18.28 | 1.55 | 0.33 | 0.04 |
| 100 | 20.25 | 12.39 | 1.14 | 0.23 | 0.03 |

**TABLE XIII: Radiation dose from photons in mSv yr$^{-1}$**

| Magnetic moment (%) | 100 gcm$^{-2}$ | 200 gcm$^{-2}$ | 500 gcm$^{-2}$ | 700 gcm$^{-2}$ | 1036 gcm$^{-2}$ |
|---|---|---|---|---|---|
| 0 | 405.20 | 203.14 | 13.21 | 2.12 | 0.40 |
| 15 | 381.99 | 203.50 | 14.02 | 2.38 | 0.24 |
| 50 | 198.59 | 112.05 | 9.29 | 1.78 | 0.22 |
| 100 | 133.49 | 77.49 | 6.74 | 1.29 | 0.18 |

Based on modeling particle fluxes and their atmospheric interactions, we found that although the magnetic field shielding is an important factor deciding the radiation dose on the surface, the atmospheric thickness is the dominating factor. If the atmosphere is sufficiently thick, such as in case of the earth, the radiation levels only increase by a factor of ~ 2 even in case of no magnetic shielding. On the other hand, the GCR induced dose increase is very large, ~ 1600, when the atmospheric thickness is ~10% that of the Earth. Comparing with the total annual natural background radiation (2.4 mSv/yr), the increase in radiation dose is by a factor of 230. Although, it is hard to assess the long-term impact of radiation dose, lethal doses calculated for Earth-based life can be taken as a reasonable upper limit. A total radiation dose of 4 Sv is considered to be lethal for humans, resulting in a 90% probability of death (United States Nuclear Regulatory Commission, 2013). A planet with 100 gcm$^{-2}$ atmosphere and less than 15% of the Earth's magnetic moment would cross this limit in less than 10 years. Such radiation is certainly not suitable for a sustained habitat for Earth-like life. In addition to the liquid water habitability criteria, biological radiation dose should also be considered as an important factor in constraining the habitability of a planet.

**TABLE XIV: Radiation dose from neutrons in mSv yr$^{-1}$**

| Magnetic moment (%) | 100 gcm$^{-2}$ | 200 gcm$^{-2}$ | 500 gcm$^{-2}$ | 700 gcm$^{-2}$ | 1036 gcm$^{-2}$ |
|---|---|---|---|---|---|
| 0 | 72.54 | 19.11 | 0.62 | 0.08 | 0.003 |
| 15 | 57.30 | 17.11 | 0.62 | 0.07 | 0.002 |
| 50 | 23.68 | 7.58 | 0.34 | 0.05 | 0.001 |
| 100 | 14.82 | 4.92 | 0.23 | 0.04 | 0.001 |

**TABLE XV: Total biological radiation dose in mSv yr$^{-1}$**

| Magnetic moment (%) | 100 gcm$^{-2}$ | 200 gcm$^{-2}$ | 500 gcm$^{-2}$ | 700 gcm$^{-2}$ | 1036 gcm$^{-2}$ |
|---|---|---|---|---|---|
| 0 | 553.33 | 262.51 | 17.22 | 3.02 | 0.65 |
| 15 | 509.81 | 260.48 | 18.20 | 3.38 | 0.46 |
| 50 | 257.21 | 141.20 | 11.98 | 2.53 | 0.42 |
| 100 | 171.46 | 97.04 | 8.70 | 1.84 | 0.34 |

## ACKNOWLEDGEMENTS

We thank Adrian Melott, Shimantini Shome and Dinesh Atri for helpful comments.

## REFERENCES

[1] Kasting, James F., Daniel P. Whitmire, and Ray T. Reynolds. "Habitable Zones around Main Sequence Stars." Icarus 101, no. 1 (1993): 108-128.

[2] Atri, Dimitra, and Adrian L. Melott. "Cosmic Rays and Terrestrial Life: A brief review." Astroparticle Physics (2013).

[3] Melott, Adrian L., and Brian C. Thomas. "Astrophysical ionizing radiation and earth: a brief review and census of intermittent intense sources." Astrobiology 11, no. 4 (2011): 343-361.

[4] Dartnell, Lewis R. "Ionizing radiation and life." Astrobiology 11, no. 6 (2011): 551-582.

[5] Ejzak, Larissa M., Adrian L. Melott, Mikhail V. Medvedev, and Brian C. Thomas. "Terrestrial consequences of spectral and temporal variability in ionizing photon events." The Astrophysical Journal 654, no. 1 (2007): 373.

[6] Thomas, Brian C., Adrian L. Melott, Charles H. Jackman, Claude M. Laird, Mikhail V. Medvedev, Richard S. Stolarski, Neil Gehrels, John K. Cannizzo, Daniel P. Hogan, and Larissa M. Ejzak. "Gamma-ray bursts and the Earth: Exploration of atmospheric, biological, climatic, and biogeochemical effects." The Astrophysical Journal 634, no. 1 (2005): 509.

[7] Gehrels, Neil, Claude M. Laird, Charles H. Jackman, John K. Cannizzo, Barbara J. Mattson, and Wan Chen. "Ozone depletion from nearby supernovae." The Astrophysical Journal 585, no. 2 (2003): 1169.

[8] Atri, Dimitra, Adrian L. Melott, and Brian C. Thomas. "Lookup tables to compute high energy cosmic ray induced atmospheric ionization and changes in atmospheric chemistry." Journal of Cosmology and Astroparticle Physics 2010, no. 05 (2010): 008.

[9] Atri, Dimitra, and Adrian L. Melott. "Modeling high-energy cosmic ray induced terrestrial muon flux: A lookup table." Radiation Physics and Chemistry 80, no. 6 (2011): 701-703.

[10] Overholt, Andrew, Adrian Melott, and Dimitra Atri. "Modeling high-energy cosmic ray induced terrestrial and atmospheric neutron flux: A lookup table." Journal of Geophysical Research - Space Science (2013).

[11] Grießmeier, J-M., A. Stadelmann, U. Motschmann, N. K. Belisheva, H. Lammer, and H. K. Biernat. "Cosmic ray impact on extrasolar Earth-like planets in close-in habitable zones." Astrobiology 5, no. 5 (2005): 587-603.

[12] Grießmeier, J-M., A. Stadelmann, J. L. Grenfell, H. Lammer, and U. Motschmann. "On the protection of extrasolar Earth-like planets around K/M stars against galactic cosmic rays." Icarus 199, no. 2 (2009): 526-535.

[13] Grenfell, John Lee, Jean-Mathias Grießmeier, Beate Patzer, Heike Rauer, Antigona Segura, Anja Stadelmann, Barbara Stracke, Ruth Titz, and Philip Von Paris. "Biomarker Response to Galactic Cosmic Ray-Induced NOx And The Methane Greenhouse Effect in The Atmosphere of An Earth-Like Planet Orbiting An M Dwarf Star." Astrobiology 7, no. 1 (2007): 208-221.

[14] Grenfell, John Lee, Jean-Mathias Grießmeier, Philip von Paris, A. Beate C. Patzer, Helmut Lammer, Barbara Stracke, Stefanie Gebauer, Franz Schreier, and Heike Rauer. "Response of Atmospheric Biomarkers to NO x-Induced Photochemistry Generated by Stellar Cosmic Rays for Earth-like Planets in the Habitable Zone of M Dwarf Stars." Astrobiology 12, no. 12 (2012): 1109-1122.

[15] Heck, Dieter, J. Knapp, J. N. Capdevielle, G. Schatz, and T. Thouw. "CORSIKA: A Monte Carlo code to simulate extensive air showers". Vol. 6019. FZKA, 1998.

[16] Risse, M., D. Heck, and J. Knapp. "EAS simulations at Auger energies with CORSIKA." International Cosmic Ray Conference, vol. 2, p. 522. 2001.

[17] Bernlhr, Konrad. "Simulation of imaging atmospheric Cherenkov telescopes with CORSIKA and sim−telarray." Astroparticle Physics 30, no. 3 (2008): 149-158.

[18] Nagano, Motohiko, D. Heck, K. Shinozaki, N. Inoue, and J. Knapp. "Comparison of AGASA data with CORSIKA simulation." Astroparticle Physics 13, no. 4 (2000): 277-294.

[19] Scalo, John, and J. Craig Wheeler. "Astrophysical and astrobiological implications of gamma-ray burst properties." The Astrophysical Journal 566, no. 2 (2002): 723.

[20] Voigt, Gerd-Hannes. "A mathematical magnetospheric field model with independent physical parameters." Planetary and Space Science 29, no. 1 (1981): 1-20.

[21] Stadelmann, A., J. Vogt, K-H. Glassmeier, M-B. Kallenrode, and G-H. Voigt. "Cosmic ray and solar energetic particle flux in paleomagnetospheres." Earth, Planets, and Space 62 (2010): 333-345.

[22] Gaisser, Thomas K. "Cosmic rays and particle physics". Cambridge University Press, 1991.

[23] Nicolet, Marcel. "Stratospheric ozone: An introduction to its study." Reviews of Geophysics 13, no. 5 (1975): 593-636.

[24] Ruderman, Malvin A. "Possible consequences of nearby supernova explosions for atmospheric ozone and terrestrial life. "Science 184, no. 4141 (1974): 1079-1081.

[25] Tarter, Jill C., Peter R. Backus, Rocco L. Mancinelli, Jonathan M. Aurnou, Dana E. Backman, Gibor S. Basri, Alan P. Boss et al. "A reappraisal of the habitability of planets around M dwarf stars." Astrobiology 7, no. 1 (2007): 30-65.

[26] Scalo, John, Lisa Kaltenegger, Antgona Segura, Malcolm Fridlund, Ignasi Ribas, Yu N. Kulikov, John L. Grenfell et al. "M stars as targets for terrestrial exoplanet searches and biosignature detection." Astrobiology 7, no. 1 (2007): 85-166.

[27] United States Nuclear Regulatory Commission. "Standards for protection against radiation", NRC Regulations, Title 10, Code of Federal regulations, Part 20 (2013)

[28] Francis, Z., S. Incerti, M. Karamitros, H. N. Tran, and C. Villagrasa. "Stopping power and ranges of electrons, protons and alpha particles in liquid water using the Geant4-DNA package." Nuclear Instruments and Methods in Physics Research Section B: Beam Interactions with Materials and Atoms 269, no. 20 (2011): 2307-2311.

[29] Cossairt, J. Donald, and Kamran Vaziri. "Neutron dose per fluence and weighting factors for use at high energy accelerators." Health physics 96, no. 6 (2009): 617-628.

[30] Beringer, J., J-F. Arguin, R. M. Barnett, K. Copic, O. Dahl, D. E. Groom, C-J. Lin et al. "Review of particle physics." Physical Review D 86, no. 1 (2012): 010001.

[31] Dar, Arnon, Ari Laor, and Nir J. Shaviv. "Life extinctions by cosmic ray jets." Physical review letters 80, no. 26 (1998): 5813-5816.

[32] Alpen, Edward L. "Radiation biophysics". Academic Press, 1997.

[33] Khodachenko, M. L., I. Alexeev, E. Belenkaya, H. Lammer, J-M. Grießmeier, M. Leitzinger, P. Odert, T. Zaqarashvili, and H. O. Rucker. "Magnetospheres of hot Jupiters:

The importance of magnetodiscs in shaping a magnetospheric obstacle." The Astrophysical journal 744, no. 1 (2011).